\documentclass[10pt]{iopart}
\usepackage{iopams}
\usepackage{epsf}
\usepackage{graphicx,graphics}
\begin{document}

%%%%%%
\title{Status of the joint LIGO--TAMA300 inspiral analysis}
%%%%%%

%%%Author List%%%
\author{
%%%%%%
Stephen Fairhurst$^{1}$\footnote[2]{E-mail:sfairhur@gravity.phys.uwm.edu} for
the LIGO Scientific Collaboration and 
Hirotaka Takahashi $^{2,3}$\footnote[1]{E-mail:hirotaka@vega.ess.sci.osaka-u.ac.jp}
for the TAMA Collaboration}
%Masaki Ando$^{4}$,
%Nobuyuki Kanda$^{6}$,
%Hideyuki Tagoshi$^{1}$,
%Ryutaro Takahashi$^{10}$,
%Daisuke Tatsumi$^{10}$,
%Yoshiki Tsunesada$^{10}$ and the TAMA Collaboration}

\address{
$^{1}$\ Department of Physics,
University of Wisconsin-Milwaukee, Milwaukee, WI 53201, USA}
\address{
${}^{2}$\ Department of Earth and Space Science , 
Graduate School of Science, Osaka University, 
Toyonaka, Osaka 560--0043, Japan}
\address{
${}^{3}$\ Graduate School of Science and Technology, 
Niigata University, Niigata, Niigata 950-2181, Japan}
%\address{
%$^{3}$\ Department of Physics,
%University of Wisconsin-Milwaukee, Milwaukee, WI 53201, USA}
%\address{
%${}^{4}$Department of Physics, University of Tokyo, Hongo, Bunkyo-ku, 
%Tokyo 113-0033, Japan}
%\address{
%${}^{6}$Department of Physics, Graduate School of Science, 
%Osaka City University, Sumiyoshi-ku, Osaka, Osaka 558-8585, Japan}
%\address{
%${}^{10}$National Astronomical Observatory of Japan, Mitaka, Tokyo 181-8588, 
%Japan}
%%%End_Author List%%%

%%%Abstract%%%
\begin{abstract}
%%%%%%

We present the status of the joint search for gravitational waves from
inspiraling neutron star binaries 
in the LIGO Science Run 2 and TAMA300 Data Taking Run
8 data, which was taken from February 14 to April 14, 2003, 
by the LIGO and TAMA collaborations.  
In this paper we discuss what has been learned from an analysis of a subset of the
data sample reserved as a ``playground''.  We determine the coincidence
conditions for parameters such as the coalescence time and chirp mass by
injecting simulated Galactic binary neutron star signals into the data stream.
We select coincidence conditions so as to maximize our efficiency of detecting
simulated signals.  We obtain an efficiency for our coincident search of 78\%,
and show that we are missing primarily 
very distant signals for TAMA300.
%the most distant signals for TAMA300.
We perform a time slide analysis to estimate the background due to accidental
coincidence of noise triggers.  We find that the background triggers have a
very different character from the triggers of simulated signals.

%%%%%%
\end{abstract}
%%%End_Abstract%%%

%%%%Uncomment for PACS numbers title message%%%
%%%%%%

%\maketitle
%%%%%%

%%%%%%
\section{Introduction}\label{sec:intro}
%%%%%%

In the past several years, there has been substantial progress in
gravitational wave detection experiments by the ground-based laser
interferometers, LIGO~\cite{ref:ligo}, TAMA300~\cite{ref:tama},
GEO600~\cite{ref:geo} and VIRGO~\cite{ref:virgo}. 

The LIGO and TAMA collaborations are conducting joint searches for
gravitational waves in the LIGO Science Run 2 (S2) and TAMA300 Data Taking Run
8 (DT8) data, which was taken from February 14 to April 14, 2003.  Three
classes of gravitational wave transients are being sought: unmodeled
gravitational wave bursts from the gamma ray burst event GRB
030329~\cite{ref:grb}; unmodeled gravitational wave bursts without an
electromagnetic trigger~\cite{ref:burst_ligotama}; inspiral signals from
Galactic binary neutron star (BNS) systems.

In this paper, we present the status of the joint search for gravitational
waves from inspiraling neutron star binaries by the LIGO and TAMA
collaborations.  We focus on what has been learned from a subset (10\%) of the
full coincident data set, chosen to be representative of the full data set in
term of detector contribution and noise, and reserved for tuning of the
analysis pipeline.  We refer to this as the playground data. 

We use the observable effective distance to characterize the sensitivity of the
instruments to binary neutron star inspirals.   This is defined as the
distance at which an inspiral of $1.4$--$1.4 M_{\odot}$ neutron stars, in the
optimal direction and orientation with respect to each detector, would produce
a signal to noise ratio (SNR) of 8.  The effective distance is always greater
than or equal to the actual distance, and on average is $2.3$ times as large
as the distance.  In Table~\ref{tab:obseve_dist}, we show the observable
effective distance of the three LIGO instruments (H1, H2 and L1) and the
TAMA300 instrument (T1) during S2 and DT8.  Since the LIGO and TAMA300
detectors were sensitive to the majority of the Milky Way, it is possible for
us to detect an inspiral signal from our Galaxy.  If we do not make a
detection, then we can place an upper limit on the binary neutron star
inspiral rate in the Milky Way.  Another motivation for this work is to gain
experience in performing coincident analyses between different groups, and to
establish the analysis method. 

\begin{table}[htpd]
\begin{center}
\begin{tabular}{c c c}
\hline \hline
Detector & Observable effective distance & Comments\\
\hline
 L1 & $\sim 2000$ kpc & Reach M31, M32, M33, M110\\
 H1 & $\sim  900$ kpc & Barely reach M31 etc  \\
 H2 & $\sim  600$ kpc& Covered the Milky Way. \\
 T1 & $\sim   50$ kpc& Covered most of the Milky Way. \\
 \hline\hline
\end{tabular} 
\end{center} 
\caption{The observable effective distance, during the S2 and DT8, at which an
inspiral of 1.4-1.4 $M_{\odot}$ neutron stars, in the optimal direction and
orientation with respect to each detector, would produce the SNR of 8.}  
\label{tab:obseve_dist} 
\end{table}

Each group analyzes its own data, generating a ``trigger'' when an inspiral
candidate appears in the data stream.  The groups then exchange triggers and
search for coincidences.  Both LIGO and TAMA300 have performed independent
analyses on the S2/DT8 data set and they have been described in detail
elsewhere \cite{ref:tama_DT8, ref:LIGO_S2_ins}.  In this paper, we will
focus on aspects of the coincident search. Using the playground data, we
define a set of coincidence requirements between triggers from the two
detectors.  Before we perform the coincident analysis using these coincidence
conditions, they are tested by performing software injections of simulated
inspiral signals into the data and by checking the detection efficiency.  We
verify that the injection efficiency is consistent with the observable
effective distances stated in Table~\ref{tab:obseve_dist}. Finally, using a
technique of artificially shifting the time series with respect to one another,
we estimate the coincident triggers produced accidentally by the instrumental noise.

This paper is organized as follows. In section~\ref{sec:detector_data_set}, we
describe the LIGO and TAMA300 detectors and the data sets to be analyzed.  In
section \ref{sec:analysis_playground}, we briefly describe the single
instrument playground analysis.
%In section \ref{sec:pip}, we  briefly describe a method of matched filtering
%pipeline used for LIGO and TAMA300. 
In section \ref{sec:condition}, we discuss the method of coincident analysis
using the results of the single-detector searches.  
%In section \ref{sec:each}, the results of the matched filtering pipeline for
%each detector are shown.  
In section \ref{sec:back_etc}, we estimate the background of coincident
triggers using a time slide analysis and test the efficiency of our search
using simulations.  Section~\ref{sec:summary} summarizes the paper.

%%%%%%
\section{Detectors and data sets}\label{sec:detector_data_set}
%%%%%%

The LIGO network of detectors consists of a 4km interferometer ``L1''in
Livingston, LA and a 4km ``H1'' and a 2km ``H2'' interferometer which share a
common vacuum system in Hanford, WA. TAMA300 is a 300m interferometer ``T1''
in Mitaka, Tokyo.  Basic information on the position, orientation of these
detectors and detailed descriptions of their operation can be found
in~\cite{ref:ligo, ref:tama}.

The data analyzed in this search was taken during LIGO S2, TAMA300 DT8, between
16:00 UTC 14 February 2003 and 16:00 UTC 14 April 2003.  We only analyze data
from the periods when both LIGO and TAMA300 interferometers were operating.
Furthermore, we restrict to times when only one of the LIGO sites was
operational, thus sharing no common observational time with the LIGO only
search discussed in \cite{ref:LIGO_S2_ins}.  Therefore, we have four
independent data sets to analyze: the data set during which neither H1 nor H2
were operating, the nH1-nH2-L1-T1 coincident data set (here ``n'' stands for
``not operating''); and three data sets when one or both of the Hanford
detectors were operational but L1 was not ---  the H1-nH2-nL1-T1 coincident
data set; the nH1-H2-nL1-T1 coincident data set; and the H1-H2-nL1-T1
coincident data set.  The total length of data was 650 hours.  This comprised
334 hours of H1-H2-T1 data, 212 hours of H1-T1 data, 68 hours of H2-T1 data and
36 hours of L1-T1 data.  To avoid any bias from tuning our pipeline using the
same data from which we derive our upper limits, the tuning of analysis
parameters was done without examining the full coincident trigger sets.
Instead, preliminary tuning was done on the playground data, which will not be
used placing the upper limit.  In this analysis, the length of playground data
was 64 hours.

%%%%%%
\section{Overview of the single interferometer analysis}\label{sec:analysis_playground}
%%%%%%

Our analysis methodology is similar, though not identical, to that used in the
LIGO analysis \cite{ref:LIGO_S2_ins, ref:LIGO_S1_ins} and the TAMA300-LISM
coincident analysis \cite{ref:tama-lism}.  We analyze the data from each
detector independently, identifying candidate inspiral events in the data
streams which we refer to as ``triggers''.  We then search for coincidences
within the lists of triggers from each interferometer.  Here, we will only
briefly review the main steps in the single instrument analyses.  The search
methods employed by TAMA300 are discussed in greater detail in
\cite{ref:tama_DT8, ref:tama-lism} and by LIGO in \cite{ref:LIGO_S2_ins,
ref:LIGO_inspiral_GWDAW}. 

For each instrument we produce a bank of inspiral templates with the 2PN
expansion for LIGO~\cite{ref:Owen_and_Sathya} and the 2.5PN expansion for TAMA300
\cite{ref:Tanaka_and_Tagoshi}, distinguished by
the masses ($m_{1}$ and $m_{2}$) of the two neutron stars in the binary.  We
then filtered the data stream through this bank of templates and record a
trigger whenever the SNR exceeds 7.  In order to minimise the number of
spurious triggers, we also perform a test of waveform consistency (the
$\chi^{2}$ test~\cite{ref:Allen}).  Additionally, during times when both
Hanford detectors are operational, we check for coincidence between H1 and H2.
In the playground data, there were no such triggers.  However, if a trigger is
seen in H1 and its amplitude, or estimated effective distance, is such that we
would not expect to see it in H2, we keep this H1 trigger even with no
coincident trigger in H2 (see \cite{ref:LIGO_S2_ins} for further details). 

The triggers from each instrument are characterized by the time of coalescence
$t_c$, the chirp mass ${\cal M} \equiv M \eta^{3/5}$,  the symmetric mass ratio
(dimensionless) $\eta \equiv m_1 m_2 / M^2$, ($M = m_{1} + m_{2}$ is the total mass
of the binary system), the signal to noise ratio $\rho$, value of the $\chi^2$ test and
effective distance $D_{\rm {eff}}$.  The effective distance of a waveform is
the distance for which a binary neutron star system would produce the
waveform if it were optimally oriented.  The lists of triggers are then
exchanged between the LIGO and TAMA analysis groups and coincident triggers
are searched for. 

%%%%%%
%\section{Threshold for the triggers}\label{sec:pip}
%%%%%%
%
%In this section, we describe the method of matched filtering pipeline
%in each detector. 
%This was independently done by LIGO and TAMA300 using different algorithms 
%and threshold.
%The algorithms of LIGO pipeline is described in~\cite{ref:LIGO_S1_ins}
%and the algorithms of the TAMA300 pipeline is described in \cite{ref:tama-lism}.
%
%The LIGO analysis pipeline~\cite{ref:LIGO_S1_ins} 
%was run on the playground data. 
%We imposed the signal to noise ratio threshold $\rho_m$ and
%$\chi^2$ threshold as :
%\begin{equation}
%\rho_m > 7, \ \ \chi^2 < 5.0 \{ 15 + (0.15 \rho)^2 \}.
%\end{equation}
%In LIGO analysis pipeline, the degrees of freedom of the $\chi^2$ is 28.
%We will look at the triggers produced in the four distinct LIGO times: 
%L1 only, H1 only, H2 only and H1-H2 coincident time.
%
%The TAMA300 analysis pipeline~\cite{ref:tama-lism} was applied to the playground data. 
%We imposed the signal to noise ratio threshold $\rho_m$ and
%$\chi^2$ threshold as :
%\begin{equation}
 %\rho_m > 7, \ \ \chi^2 < 2.4 \{ 16 + (0.185 \rho)^2 \}.
%\end{equation}
%In TAMA analysis pipeline, the degrees of freedom of the $\chi^2$ is 30.
%Note that
%we introduce a $\chi^2$ cut in the way described in \cite{ref:LIGO_S1_ins} and~\cite{ref:grasp}. 
%This is different from the $\chi^2$ cut of TAMA300 analysis (see~\cite{ref:tama-lism}
%and~\cite{ref:tama_DT8}). 
%This is because the coincident analysis will become much simpler 
%if LIGO and TAMA use the same criterion for $\chi^2$. 

%%%%%%
\section{Tuning of the coincident analysis}\label{sec:condition}
%%%%%%

True gravitational wave events, if sufficiently loud, will appear in the
trigger lists of both the LIGO and TAMA300 instruments.  However, they will
have somewhat different values of parameters such as coalescence time and
chirp mass.  This is due to the detectors' noise, the difference in the
detectors' locations and arm orientations, and the discrete sampling of the
mass parameter space by the template bank.  To evaluate the accuracy with
which we can determine various parameters, we perform Monte Carlo simulations
of injected signals.  Our sample population for simulations consisted of
Galactic neutron star binaries with masses in the range $1.0 M_{\odot} \le
m_1,m_2 \le 3.0 M_{\odot}$.  For the simulations, the spatial and mass
distributions were taken from \cite{ref:Belczynski} and \cite{ref:Kim}.
%whilefor TAMA the distributions given in \cite{ref:Curran_and_Lorimer} were used.
We set coincidence windows to maximize our detection efficiency while reducing
the number of chance coincidences.  In the LIGO--TAMA300 search we require
coincidence in both time and chirp mass.

%%%%%%
\subsection{Timing accuracy} 
%%%%%%

\begin{figure}[htpd]
\centering
\scalebox{0.4}[0.4]{\includegraphics{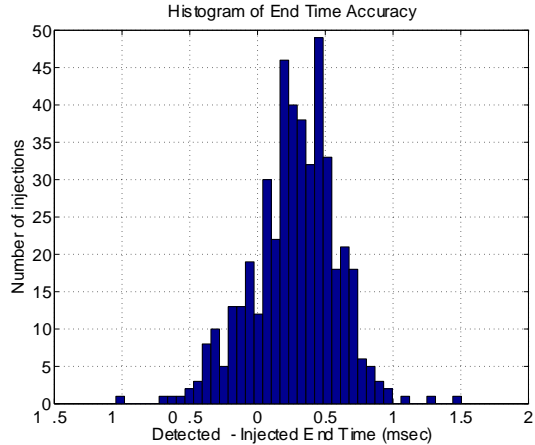}}
\caption{Accuracy of the reconstructed coalescence time $t_c$ for simulated
injections into LIGO data.  All of the 455 injected signals were seen
($\rho>7$) in the LIGO data, and the coalescence time of the majority of them
was recorded within $\pm1.0$ ms of the injection coalescence time.}
\label{fig:L_acc_tc}
\end{figure}

\begin{figure}[htpd]
\centering
\scalebox{0.4}[0.4]{\includegraphics{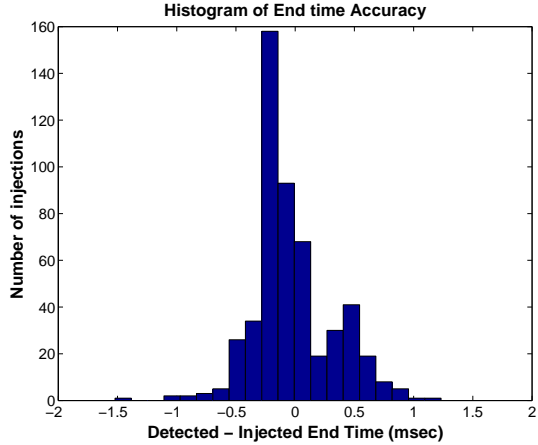}}
\caption{Accuracy of $t_c$ for simulated injections into TAMA300 data.  Out of
the 660 injections into TAMA300 data, 516 produced triggers  ($\rho>7$).  For
the majority of these injections, the coalescence time was recorded within
$\pm 1.0$ ms of the injection time. The other simulated signals injected into
TAMA300 data did not produce triggers because they were too weak to be
detected by TAMA300.}
\label{fig:T_acc_tc}
\end{figure}

We require the coalescence time $t_c$ of a signal to be consistent between the
LIGO and TAMA300 detectors.  The allowed time difference is determined by our
accuracy in determining $t_c$ and the light travel time between the sites. To
determine our timing accuracy,  we inject simulated gravitational wave signals
into the data streams from each detector, and re-analyze the data.
%in exactly the same manner as is done in the actual gravitational wave search.
%These simulations require that we determine a target population, specified by
%the wave form and the distribution of sources over the sky.  The injection
%time is determined so that there is one injection in each of the 800
%playground segments.  Out of the 800 injection time, 660 are in the
%observation time of TAMA300 455 are in the observation time of LIGO. These
%time are used as the actual injection time. 
We injected 660 signals in TAMA300 data and 455 signals in LIGO data.  We
compared the injected time and the detected time which was obtained from our
search pipeline.  We define the detected time to be the time of the loudest
trigger within 3 ms of the coalescence time of the injection.  We show the
accuracy of coalescence time; detected - injected coalescence time in
Figs.~\ref{fig:L_acc_tc} and \ref{fig:T_acc_tc}.  We find that nearly all
injections into both LIGO and TAMA300 data have triggers ($\rho>7$) recorded
within $\pm1.0$ ms of the injection time.

%All of the 455 injections into LIGO data
%have triggers ($\rho>7$) recorded within $\pm1.5$ ms of the injection time.
%Out of the 660 injections into TAMA300 data,
%516 have triggers  ($\rho>7$)  recorded
%within $\pm 1.5$ ms of the injection time. 
%The other simulated signals injected into TAMA300 data did not produce the triggers
%because they were too weak to be detected by TAMA300.

The distance between the TAMA300 site and the LIGO Hanford site is 7487 km.
The distance between the TAMA300 site and the LIGO Livingston site is 9703 km.
Because gravitational waves travel at the speed of light, the maximum delay of
the arrival time of gravitational wave signals are $\Delta t_{{\rm dist
H-T}}=25.0 \mathrm{ms}$ for TAMA300--Hanford and $\Delta t_{{\rm dist
L-T}}=32.4 \mathrm{ms}$ for TAMA300--Livingston.  Based on these facts, we
define the allowed difference in coalescence time $\Delta t_{{\rm w}}$ by the
maximum time delay of signal plus the uncertainty in determining the
coalescence time.  The values are
\begin{eqnarray}
  \Delta t_{{\rm w\ H-T}} &=& 27 \mbox{ms for Hanford--TAMA300 and} 
  \nonumber \\
  \Delta t_{{\rm w\ L-T}} &=& 35 \mbox{ms for Livingston--TAMA300} \, .
\end{eqnarray}
Therefore, if the coalescence times, $t_{c, {\rm T1}}$ for TAMA300 and $t_{c,
i}$ (where $i=$ H or L) for LIGO, of a trigger satisfy,
\begin{equation} 
  | t_{c, {\rm T1}} - t_{c, i}| < \Delta t_{{\rm w}\ i {\rm-T}}, 
\end{equation} 
the trigger is recorded as a candidate event.

%%%%%%
\subsection{Chirp mass accuracy}
%%%%%%

\begin{figure}[htpd]
\centering
\scalebox{0.4}[0.4]{\includegraphics{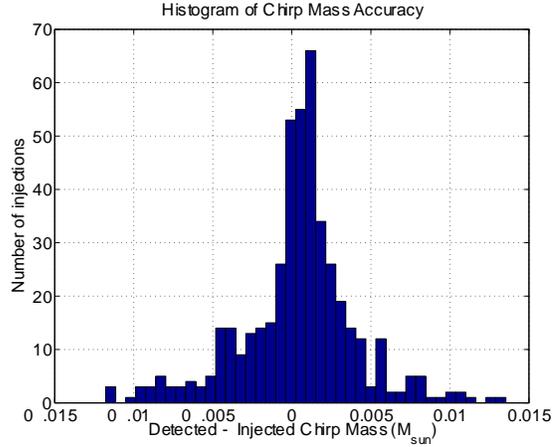}}
\caption{Accuracy of chirp mass $\mathcal{M}$ for simulated injections into
LIGO data.  The majority of the 455 injections into LIGO data have triggers
($\rho>7$) recorded with chirp mass within $0.01M_{\odot}$ of the chirp mass
of the injected signal. }
\label{fig:L_acc_mass}
\end{figure}

\begin{figure}[htpd]
\centering
\scalebox{0.4}[0.4]{\includegraphics{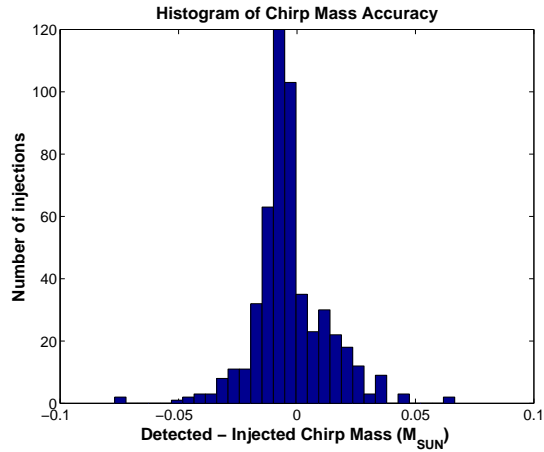}}
\caption{Accuracy of chirp mass $\mathcal{M}$ for simulated injections into
TAMA300 data.  Out of the 516 found injections, the majority have triggers
($\rho>7$) recorded with chirp mass within $0.05M_{\odot}$ of the chirp mass
of injected signal. }
\label{fig:T_acc_mass}
\end{figure}

%We have learned in TAMA-LISM coincident analysis~\cite{ref:tama-lism} that
%the error in measuring the reduced mass was so large that it did not affect
%the coincident analysis very much.  Thus, we only take into account of the
%chirp mass difference here. 
The accuracy with which we can determine the chirp mass is determined in the
same way as for the coalescence time; namely by adding simulated signals into
the data of the instruments.  As with the end time, we take the detected chirp
mass to be the chirp mass of the loudest trigger obtained within 3 ms of the
end of the injection. In Figs.~\ref{fig:L_acc_mass} and \ref{fig:T_acc_mass},
we find that the chirp mass of an injected signal is recorded with an accuracy
of $0.01M_{\odot}$ in the LIGO instruments and $0.05M_{\odot}$ in TAMA300.  We
define the allowed difference in chirp mass of coincident triggers between
LIGO and TAMA300 to be:

\begin{equation}
  \Delta \mathcal{M}_{{\rm w}}=0.05M_\odot \, .
\end{equation}  
If the chirp mass, $\mathcal{M}_{{\rm T1}}$ for TAMA300 and $\mathcal{M}_{i}$
(where $i=$ H or L) for LIGO, of a trigger satisfies, 
\begin{equation} |
\mathcal{M}_{{\rm T1}} - \mathcal{M}_{i}| < \Delta \mathcal{M}_{{\rm w}},
\end{equation}
the trigger is recorded as a candidate event.

\section{Background, efficiency and consistency check of the analysis}
\label{sec:back_etc}
%%%%%%

Before doing the coincidence search, we study the background events by
performing a time slide analysis of the data.  Additionally, we evaluate the
search efficiency by performing an injection analysis.  We check that the
injection efficiency is consistent with the sensitivities of the instruments
reported in Table~\ref{tab:obseve_dist}.

%%%%%%
\subsection{Background}
 %%%%%%

\begin{figure}
\centering
\scalebox{0.4}[0.4]{\includegraphics{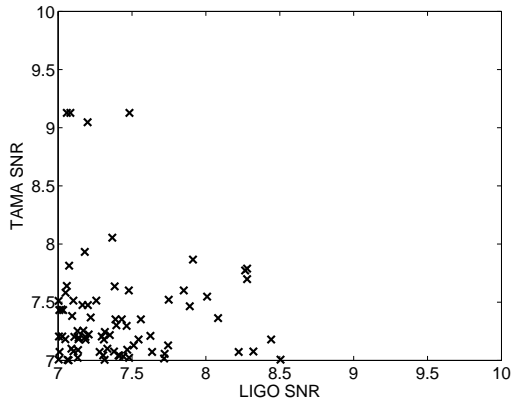}}
\caption{The signal to noise ratios $\rho_{\mbox{\rm\tiny LIGO}}$ vs
$\rho_{\mbox{\rm\tiny TAMA}}$ of the accidental coincident triggers, in
playground data, using 100 time slides.}
\label{fig:scatter_acc_coin}
\end{figure}

Even in the absence of gravitational wave signals, we expect that some noise
triggers will be coincident by chance.  We can estimate the number of such
accidental coincident triggers by performing our coincidence analysis on the
data with artificial relative time shifts added~\cite{ref:Amaldi_Astone}.  We
performed 100 time slides of $\pm 5, 10, 15, 20, \ldots, 250$ sec of the LIGO
triggers relative to the TAMA300 triggers. These slides are much longer than
the light travel time between the sites, so that any coincidence cannot be
from actual gravitational wave.  They are also longer than the detector noise
auto-correlation time, longer than the longest signal template duration and
shorter than timescales of detectors' non-stationarity, so that each time
slide provides an independent estimate of the accidental coincident rate.
Additionally, the triggers are clustered over a 100 ms window.  The signal to
noise ratios $\rho_{\mbox{\rm\tiny LIGO}}$ vs $\rho_{\mbox{\rm\tiny TAMA}}$ of
the accidental coincident triggers found in the playground data is shown in
Fig~\ref{fig:scatter_acc_coin}.  From these time slides, we obtain an
estimated false alarm rate of 0.4 triggers per day.

%%%%%%
\subsection{Efficiency and consistency check of the analysis}
%%%%%%

We performed a set of coherent injections into the data of both TAMA300 and
the LIGO instruments in order to test our coincidence windows and determine
the search efficiency.  Overall, this consisted of 381 injections in times
when TAMA300 and one of the LIGO sites were operational.
These injections were same as  those described in Section \ref{sec:condition} 
used for determining the coincidence parameters.
%These injections
%were distinct from those described in Section \ref{sec:condition} used for
%determining the coincidence parameters, although the spatial and mass
%distributions followed those used in the LIGO injections.  

Among the 381 injections, 297 events produced coincident triggers in the two
instruments.  Of the 84 injections which were not found by the coincidence
analysis, only one injection was found separately by both LIGO and TAMA300.
However, in TAMA300, the chirp mass was recovered with an error of $0.07
M_{\odot}$, while LIGO recovered it with an error of less than $0.01
M_{\odot}$.  Hence, although a trigger was found in both instruments, it did
not pass the chirp mass coincidence requirement.

\begin{figure}
\centering
\scalebox{0.4}[0.4]{\includegraphics{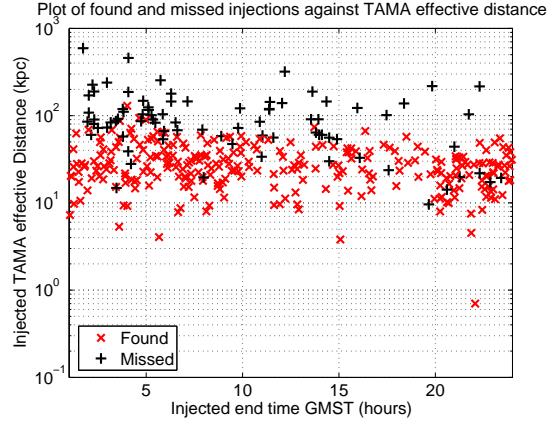}}
\caption{Injected distance versus GMST for Galactic BNS inspiral signals
injected into the TAMA300 playground data stream. This figure shows injected
signals which were found by the LIGO-TAMA300 inspiral pipeline ($\times$), and
those which were missed ($+$).  The majority of missed injections are at
distances greater than the range of the TAMA300 instrument during DT8.}
\label{fig:distance}
\end{figure}

\begin{figure}
\centering
\scalebox{0.4}[0.4]{\includegraphics{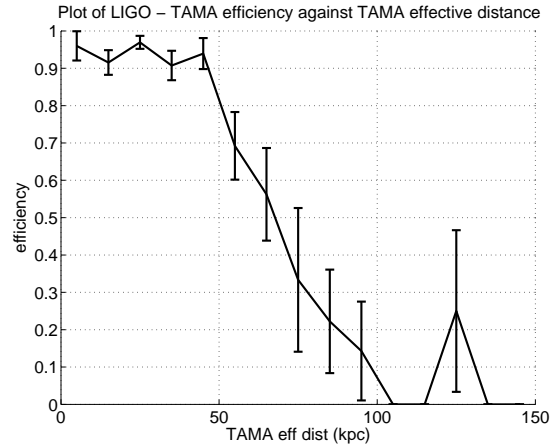}}
\caption{LIGO--TAMA300 coincident detection efficiency for Galactic BNS inspirals
versus effective distance of the signal at the TAMA300 detector.}
\label{fig:eff_hist}
\end{figure}

The other 83 missed injection signals were not detected by TAMA300 with $\rho
> 7$ and $\chi^{2}$ below threshold.  We have checked that the missed
injections are actually very distant signals for TAMA300.  To do this, we plot
the effective distance versus GMST for the found and missed injections in
Fig.~\ref{fig:distance}.  It is clear that the majority of missed injections
are too distant for TAMA300 to detect.  We also plot the efficiency versus
effective distance in Fig.\ref{fig:eff_hist}.  We find that the efficiency is
very close to 100\% within 50 kpc and then drops off steeply.  This is
consistent with the observable distance of TAMA300 given in
Table~\ref{tab:obseve_dist}. 

\begin{figure}
\centering
\scalebox{0.4}[0.4]{\includegraphics{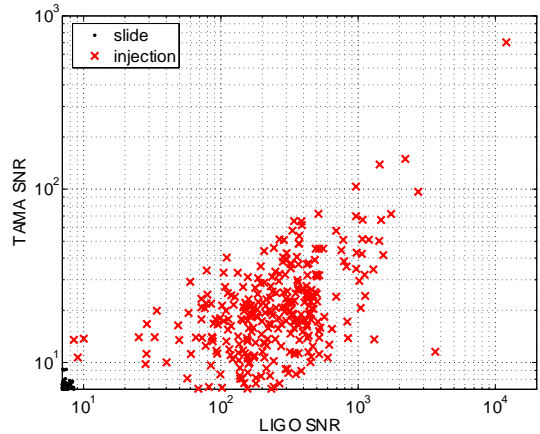}}
\caption{The signal to noise ratios of inspiral triggers from LIGO versus
TAMA300, including the time slide triggers ($\bullet$) and the playground
injection triggers ($\times$).}
\label{time_slide_and_injected}
\end{figure}

In Fig.~\ref{time_slide_and_injected}, we show the scatter plot
($\rho_{\mbox{\rm\tiny LIGO}}, \rho_{\mbox{\rm\tiny TAMA}}$) with both the
time slide and injection triggers from the playground analysis. The plot shows
clearly that the background triggers are well separated from injections of
Galactic BNS inspiral signals making it easy to distinguish the latter from
the former.  Therefore, it would be possible for us to distinguish a Galactic
inspiral signal from the background noise. 

%%%%%%
\section{Summary}\label{sec:summary}
%%%%%%

We presented the status of the joint search for gravitational waves from
inspiraling neutron star binaries by the LIGO and TAMA collaborations.  We
focused primarily on what has been learned from playground data.

We discussed a trigger based method of performing the coincidence analysis in
a multi detector search for gravitational waves from inspiraling neutron star
binaries.  We determined efficient coincidence conditions on the coalescence
time and chirp mass by analyzing Galactic binary neutron star signals injected
into the data streams of the LIGO and TAMA300 detectors.  The choice of these
coincidence parameters was validated by performing a set of coherent simulated
injections into both LIGO and TAMA300 playground data.  Using these injections,
we found the detection efficiency to Galactic BNS inspirals for our coincident
search to be 78\%.  This value was consistent with expectations based on the
sensitivities of the instruments during S2/DT8, derived from their typical
noise spectra.  Additionally, we performed a time slide analysis of the
playground data.  This allowed us to estimate our expected background due to
chance coincidences.  Furthermore, we saw that the character of the injected
signals was very different from that of the background triggers.

The coincident analysis of all of the LIGO S2--TAMA300 DT8 coincident data will be
performed using the methods discussed in this paper.  Complete results of the
analysis which will include any detection candidates as well as an upper limit
on the event rate obtained from all of the LIGO S2--TAMA300 DT8 coincident data
will be reported in a subsequent publication.

%%%%%%
\section*{Acknowledgments}
%%%%%%
This work was supported in part by the Grant-in-Aid for Scientific Research on
Priority Areas (415) of the Ministry of Education, Culture, Sports, Science and
Technology of Japan, and in part by JSPS Grant-in-Aid for Scientific Research
Nos.~14047214 and 12640269. 

The authors gratefully acknowledge the support of the United States National
Science Foundation for the construction and operation of the LIGO Laboratory
and the Particle Physics and Astronomy Research Council of the United Kingdom,
the Max-Planck-Society and the State of Niedersachsen/Germany for support of
the construction and operation of the GEO600 detector. The authors also
gratefully acknowledge the support of the research by these agencies and by the
Australian Research Council, the Natural Sciences and Engineering Research
Council of Canada, the Council of Scientific and Industrial Research of India,
the Department of Science and Technology of India, the Spanish Ministerio de
Educacion y Ciencia, the John Simon Guggenheim Foundation, the Leverhulme
Trust, the David and Lucile Packard Foundation, the Research Corporation, and
the Alfred P. Sloan Foundation. 

%%%%%%
  
%%%%%%

\end{document}